\documentclass[pra,aps,preprint,doublespace]{revtex4}

\usepackage{amssymb}
\usepackage{amsmath}
\usepackage{epsfig}
\begin{document}

\title{Depolarization channels with zero-bandwidth noises}

\date{\today}

\author{Andrzej Dragan}
\email{Andrzej.Dragan@fuw.edu.pl} \affiliation{Instytut Fizyki
Teoretycznej, Uniwersytet Warszawski, Ho\.{z}a 69, PL-00-681
Warszawa, Poland}
\author{Krzysztof W\'odkiewicz}
\email{wodkiew@fuw.edu.pl}  \affiliation{Instytut Fizyki
Teoretycznej, Uniwersytet Warszawski, Warszawa 00--681, Poland}
\affiliation{Department of Physics and Astronomy, University of
New Mexico, Albuquerque, NM~87131-1156, USA}

\date{\today}

\begin{abstract}
A  simple model describing  depolarization channels with
zero-bandwidth  environment is presented and exactly solved. The
environment is modelled by Lorentzian, telegraphic and Gaussian
zero-bandwidth noises. Such channels can go beyond the standard
Markov dynamics and therefore can illustrate the influence of
memory effects of the noisy communication channel on the
transmitted information. To quantify the disturbance of quantum
states the entanglement fidelity between  arbitrary input and
output states is investigated.
\end{abstract}

\maketitle

\section{Introduction}

One of the most important features concerning quantum
communication is the capacity or the fidelity of quantum
information transmitted in noisy quantum channels
\cite{BennettShor}. The key factor limiting the possibilities of
communication using quantum states, is an environment-induced
noise. Uncontrolled interaction between the environment ($E$), and
the transmitted quantum state can essentially affect the state and
in consequence lower the communication capacity of the information
channel \cite{Benett97,Bruss2000}.


For qubits, a well known class of quantum noisy channels consists
of depolarizing channels \cite{ref-Nielsen}. Input information  of
such channels is stored in a density operator
$\hat{\varrho}_{\mbox{\scriptsize in}}$. Such channels can be
characterized by a probability $p$
 that  the quantum information is
distorted, and with a probability $1-p$ that the information
remains intact. In the simplest case of a single qubit transmitted
through the noisy channel the influence of noise is usually
decomposed into three interaction channels. Bit error channel
$\hat{\sigma}_x$ flipping the values of bits: $|0\rangle\mapsto
|1\rangle$, $|1\rangle\mapsto |0\rangle$; phase error channel
$\hat{\sigma}_z$ flipping the phase: $|0\rangle\mapsto |0\rangle$,
$|1\rangle\mapsto -|1\rangle$; and phase and bit error channel
$\hat{\sigma}_y$ flipping both: $|0\rangle\mapsto i|1\rangle$,
$|1\rangle\mapsto -i|0\rangle$.

The influence of these interaction channels can be written as an
incoherent combination of three unbiased terms generating bit flip
errors and phase flip errors in the form given by
\cite{ref-Preskill}:

\begin{eqnarray}
\label{unbchannel} \hat{\varrho}_{\mbox{\scriptsize out}} &=&
(1-p)\hat{\varrho}_{\mbox{\scriptsize in}}
\nonumber\\
&+&\frac{p}{3\,}\left(\hat{\sigma}_x
\hat{\varrho}_{\mbox{\scriptsize in}} \hat{\sigma}_x
+\hat{\sigma}_y \hat{\varrho}_{\mbox{\scriptsize in}}
\hat{\sigma}_y +\hat{\sigma}_z \hat{\varrho}_{\mbox{\scriptsize
in}} \hat{\sigma}_z \right).
\end{eqnarray}
This depolarizing channel is just an example of a general quantum
channel characterized by a trace-preserving, linear map
$\Phi:\hat{\varrho}_{\mbox{\scriptsize in}} \mapsto
\hat{\varrho}_{\mbox{\scriptsize out}}$. The influence of the
environment on the quantum state $\hat{\varrho}_{\mbox{\scriptsize
in}}$  can be represented in terms of a Kraus decomposition:
$\Phi(\hat{\varrho}_{\mbox{\scriptsize in}})=\sum_r \hat{K}_r
\hat{\varrho}_{\mbox{\scriptsize in}} \hat{K}_r^\dagger$, where
$\sum_r \hat{K}_r^\dagger \hat{K}_r=\hat{\openone}$
\cite{kraus1983}.

The  unbiased depolarizing channel given by
Eq.~(\ref{unbchannel}), corresponds to $\hat{K}_0 =
\sqrt{1-p}\,\hat{\openone}$, $\hat{K}_1
=\sqrt{\frac{p}{3}}\,\hat{\sigma}_x$, $\hat{K}_2
=\sqrt{\frac{p}{3}}\,\hat{\sigma}_y$ and $\hat{K}_3
=\sqrt{\frac{p}{3}}\,\hat{\sigma}_z$.

It is not difficult to derive the expression  corresponding to
Eq.~(\ref{unbchannel}), from a unitary  evolution involving an
extended Hilbert space $\mathcal{H}\otimes\mathcal{H}_E$, of the
input qubit and an additional qubit ($E$), consisting  the
environment degree of freedom. Although mathematically correct,
such a derivation has no simple or direct  physical realization in
terms of a realistic  noise. In fact in most known cases, the
environment  is much more complex, and the additional fact that
the three interaction channels corresponding to  bit error, flip
error and phase error  do not commute, leaves the incoherent
addition of these channels in the formula (\ref{unbchannel})
questionable.

A more realistic approach to  depolarization channels with bit
errors requires a better understanding of the physics involved in
the system-environment interactions. We shall assume that the
system-environment interaction is characterized by a model
Hamiltonian $H$. A unitary evolution of this combined system leads
to a Kraus decomposition described  by a time dependent map
$\Phi_t:\hat{\varrho}_{\mbox{\scriptsize in}} \mapsto
\hat{\varrho}(t)$. This means that one should have a
time-dependent $p(t)$ such that at the input time $t=0$, $p(0)=0$.
Because of this the general property of the map: $\Phi_t|_{t=0} =
\hat{\openone}$, expresses the continuity at the origin and hence
for all time.

In general the problem of finding the evolution of the state
interacting with the environment is very difficult, therefore the
environment-induced noise is modelled using various simplified
approaches including several assumptions such as the Markov
property. In such case it is assumed that the evolution of the
state at a given instant is fully determined by the state at that
instant, so the process has no ``memory" of its past. The Markov
property means the quantum channel is such that for an
infinitesimal time interval: $p(\Delta t) \simeq \Delta t$. In
this case  the channel map generates a completely positive
dynamical semigroup: $\Phi_t \circ \Phi_{t^{\prime}} =
\Phi_{t+t^{\prime}}$ for $ t,t^{\prime}\geq 0$, which defines a
Markovian dynamics.

As it is well known, for such Markovian maps we can transform the
time-dependent Kraus decomposition (\ref{unbchannel}) into a local
Lindblad equation \cite{alicki1987}

\begin{equation}
\label{lindblad}
 \frac{\mbox{d}\hat{\varrho}(t)}{\mbox{d}t}= \hat{\mathcal{L}}
\hat{\varrho}(t),
\end{equation}
with the initial  condition $\hat{\varrho}(0)
=\hat{\varrho}_{\mbox{\scriptsize in}}$, and where the Lindblad
superoperator $\hat{\mathcal{L}}$ can be derived from the Kraus
operators.

In this paper we present a simple model describing an evolution of
a quantum state interacting with the environment and study various
properties of the affected state. A physical picture behind the
algebra can be for example a randomly fluctuating magnetic field
acting on an electron's magnetic moment, or a thermally
fluctuating birefringence of a single mode fiber transmitting a
polarization state of a single photon.

Using our model we will justify the assumption (\ref{unbchannel})
and determine the conditions for its validity. We will show that
the disturbed output state has the form (\ref{unbchannel}) in the
infinite interaction time limit only if the disturbance is
unbiased and acts identically in all bases. In the general case
the evolution leads to a different final state. It will also turn
out that the simple model leads to a nontrivial evolution not
obeying the Markov property. Therefore, our simple approach will
lead us out of the no-memory approximation regime. Our model shows
that in many cases one cannot fairly neglect the memory effects of
the interaction and consequently the Markov property is not always
valid \cite{ref-Goychuk,ref-Daffer}.

The paper is organized as follows: in Sec.~\ref{sec-model} we
present the model of a qubit in a presence of an external noise.
In Sec.~\ref{sec-average} we show that if the environment of the
channel consists of zero-bandwidth noises one can calculate
exactly expressions determining the evolution of a single qubit in
such channel. Sec.~\ref{sec-examples} contains several examples of
zero-bandwidth noises that can affect the input state. We show
that in general the decoherence channel can not be regarded as an
incoherent superposition of independent interactions. Special
cases involving Markovian  and exactly soluble non-Markovian
dynamics are derived. In Sec.\ref{sec-fidelity} the efficacy of
the quantum channel is quantified by the fidelity between the
output state  and the input state. An appropriate measure for
assessing the fidelity of a mixed input state is the entanglement
fidelity \cite{ref-Schumacher}, which is the maximum fidelity of
states being purifications of the input mixed state $\rho$ and the
output state $\Phi(\rho)$. Finally Sec.~\ref{sec-conclusions}
concludes the paper.

\section{Interaction model\label{sec-model}}

The interaction of a qubit with an environment inducing random bit
errors will be described by the following Hamiltonian

\begin{equation}
\hat{H} = {\bf r}\cdot  \hat{\boldsymbol{\sigma}}=
x\hat{\sigma}_x+y\hat{\sigma}_y+z\hat{\sigma}_z,
\end{equation}
where the three components $r_i=(x,y,z)$ will be uncontrolled
"noisy" parameters characterizing the fluctuations of the
environment.

We will assume that $r_{i}$ are independent random variables. This
situation is a simplification of a general very difficult problem
with  $r_{i}(t)$ being time-dependent stochastic processes with
arbitrary autocorrelations: $\langle r_{i}(t)
r_{j}(t^{\prime})\rangle= \delta_{ij}\,\Delta_{i}(t,t^{\prime})$.
In these applications, the autocorrelation functions
$\Delta_{i}(t,t^{\prime})$, have usually a Fourier limited
spectrum with an effective bandwidth $\gamma$ characterizing the
environment noise. Even for the simplest form of the
autocorrelations, the exact solution of the full time-dependent
problem involving more that one $r_{i}(t)$ noise is not known.
However our simple noise model can illustrate several properties
of various completely positive maps.

In the proposed scenario, the evolution of a single qubit given by
the von Neumann equation has  the following form

\begin{equation}
\frac{\mbox{d}\hat{\varrho}}{\mbox{d}t} = i\sum_i r_i
\left[\hat{\sigma}_i,\hat{\varrho} \right] = \hat{{\cal
L}}_{0}\hat{\varrho}.
\end{equation}
The appearing Liouville  superoperator $\hat{{\cal L}}_{0}$
describes a unitary rotation of a qubit defined by the
coefficients $r_i$. The solution  involves a stochastic averaging
with respect to the environment. As a result it can be written in
the following compact and formal form:

\begin{equation}
\label{timeorder}
 \hat{\varrho}(t) =\langle \mbox{T}
e^{\int_{0}^{t}\mbox{\scriptsize d}s\hat{\cal L}_{0}(s)}\rangle\,
\hat{\varrho}_{\mbox{\scriptsize in}}.
\end{equation}

There is no useful  formula that can handle the chronological
time-ordering of three  Pauli matrices, and allows an exact
stochastic averaging over the environment noises. There are
however special cases when the exact solution of
Eq.~(\ref{timeorder}) can be obtained. We know of three cases.
Case one involves an arbitrary stochastic noise and only one
$r_i$. In this case the chronological ordering plays no  role. In
case two, fluctuations are Gaussian and  the bandwidth
characterizing the environment noise is infinite $\gamma=\infty$
(white noise). In this case an exact average of Eq.
(\ref{timeorder}) exists, and a Lindblad equation (\ref{lindblad})
for the channel map can be derived. Case three, the one
investigated in this paper, corresponds to arbitrary random
fluctuations of $r_i$ with the environment described by a
zero-bandwidth environment noise: $\gamma=0$. In this case all
autocorrelations $\Delta_{i}(t,t^{\prime})$, become
time-independent, the chronological product plays no role, and an
exact averaging over the environment with arbitrary statistics can
be performed.

\section{Exact solution with Zero-Bandwidth \label{sec-average}}
In order to find the evolution of a state under  a
time-independent Liouvillian $\hat{{\cal L}}_{0}$, we first find
its eigenstates:

\begin{equation}
\hat{\cal L}_{0}{\bf A}\cdot\boldsymbol{\hat{\sigma}} =
ir_iA_j[\hat{\sigma}_i,\hat{\sigma}_j]=-2 r_i A_j
\epsilon_{ijk}\hat{\sigma}_k = \lambda {\bf
A}\cdot\boldsymbol{\hat{\sigma}},
\end{equation}
hence we obtain a set of equations:

\begin{equation}
-2 r_i A_j \epsilon_{ijk} = \lambda A_k.
\end{equation}
The solutions exist only for a set of eigenvalues
$\lambda\in\{0,2ir,-2ir\}$, where $r=\sqrt{x^2+y^2+z^2}$. For this
set of eigenvalues we find the corresponding eigenvectors: for
$\lambda_0 = 0$ we have ${\bf A}_0 = {\bf r}$, and for
$\lambda_\pm=\pm 2ir$ we have ${\bf A}_\pm = (\mp iyr-xz,\pm
ixr-yz,x^2+y^2)$.

In order to determine the evolution of an arbitrary initial state

\begin{equation}\label{initial}
 \hat{\varrho}_{\mbox{\scriptsize
in}}=\frac{1}{2}(\hat{\openone}+{\bf a} \cdot
{\boldsymbol{\hat{\sigma}}}),
\end{equation}
it is helpful to decompose it into the calculated eigenvectors.
Pauli operators $\hat{\sigma}_i$ written  in the calculated
eigenbasis have the following form:

\begin{eqnarray}
\hat{\sigma}_x &=& \left( \frac{x}{r^2}{\bf A}_0 - \frac{xz ({\bf
A}_+ +{\bf A}_-)-iyr({\bf A}_+ -{\bf A}_-)}
{2r^2(x^2+y^2)}\right)\cdot\boldsymbol{\hat{\sigma}}
\nonumber \\ \nonumber \\
\hat{\sigma}_y &=& \left( \frac{y}{r^2}{\bf A}_0 - \frac{yz ({\bf
A}_+ +{\bf A}_-)+ixr({\bf A}_+ -{\bf A}_-)}
{2r^2(x^2+y^2)}\right)\cdot\boldsymbol{\hat{\sigma}}
\nonumber \\ \nonumber \\
\hat{\sigma}_z &=& \left( \frac{z}{r^2}{\bf A}_0 + \frac{{\bf A}_+
+{\bf A}_-}{2r^2}\right)\cdot\boldsymbol{\hat{\sigma}}.
\end{eqnarray}
At this point it is easy to find the action of the evolution
operator $\exp(\hat{\cal L}_{0}t)$ on the given input state. We
simply multiply the eigenvectors appearing in our decomposition by
the proper factors $\exp(\lambda t)$ with corresponding
eigenvalues $\lambda$. This yields:

\begin{widetext}

\begin{eqnarray}
\label{eq-output} e^{\hat{\cal L}_{0}t}
\hat{\varrho}_{\mbox{\scriptsize in}} &=&
\frac{1}{2}\left[\hat{\openone}+a_x \left( \frac{x}{r^2}{\bf A}_0
- \frac{xz (e^{2irt}{\bf A}_+ +e^{-2irt}{\bf
A}_-)-iyr(e^{2irt}{\bf A}_+ -e^{-2irt}{\bf A}_-)}
{2r^2(x^2+y^2)}\right)\cdot\boldsymbol{\hat{\sigma}} \right. \nonumber \\ \nonumber \\
& & +a_y \left( \frac{y}{r^2}{\bf A}_0 - \frac{yz (e^{2irt}{\bf
A}_+ +e^{-2irt}{\bf A}_-)+ixr(e^{2irt}{\bf A}_+ -e^{-2irt}{\bf
A}_-)}
{2r^2(x^2+y^2)}\right)\cdot\boldsymbol{\hat{\sigma}} \nonumber \\ \nonumber \\
& & +\left.a_z \left( \frac{z}{r^2}{\bf A}_0 + \frac{e^{2irt}{\bf
A}_+ +e^{-2irt}{\bf A}_-}{2r^2}\right) \cdot
\boldsymbol{\hat{\sigma}} \right].
\end{eqnarray}

\end{widetext}

The above formula expresses the state of the qubit evolving under
the action of the Liouvillian defined by the arbitrary vector
${\bf r}$.

In our model of the noisy channel, the interaction between the
environment and the qubit can be described by  a randomly chosen
vectors ${\bf r}$. Therefore, to model the evolution of the qubit
under the influence of the environment-induced noise we will
average  the obtained output state over all possible realizations
of the dynamics characterized by  arbitrary vectors ${\bf r}$. For
simplicity we will be interested in an averaged evolution of the
qubit with an even in ${\bf r}$ probability distribution $p({\bf
r}) = p(-{\bf r})$. In this case a non-vanishing contribution to
the averaged output state will come only from the symmetric part
of the expression (\ref{eq-output}):

\begin{eqnarray}
\left\{e^{\hat{\cal L}_0t}\hat{\varrho}_{\mbox{\scriptsize
in}}\right\}_{\mbox{\scriptsize sym}} &=& \frac{1}{2}\left[
\hat{\openone}
+a_x\hat{\sigma}_x\left(\frac{x^2}{r^2}+\frac{y^2+z^2}{r^2}\cos
2rt\right)\right.
\nonumber \\ \nonumber \\
& &+a_y\hat{\sigma}_y\left(\frac{y^2}{r^2}+\frac{x^2+z^2}{r^2}\cos
2rt\right) \nonumber \\ \nonumber \\
&
&+\left.a_z\hat{\sigma}_z\left(\frac{z^2}{r^2}+\frac{x^2+y^2}{r^2}\cos
2rt\right)\right].
\end{eqnarray}
From the above formula it follows that the initial state
(\ref{initial}) evolves into an averaged output state

\begin{equation}\label{output}
 \hat{\varrho}_{\mbox{\scriptsize out}}(t)=
\frac{1}{2}\left(\hat{\openone}+a_i\Lambda_i(t)\hat{\sigma}_i\right).
\end{equation}
The dynamics of the output state  is completely described at any
time by a set of time-dependent functions $\Lambda_i(t)$, which
have the following form:

\begin{equation}
\Lambda_i(t) = 1-2\int \mbox{d}^3r\,p({\bf
r})\left(1-\frac{r_i^2}{r^2}\right)\sin^2 rt.
\end{equation}
The output state can be also equivalently represented in terms of
Kraus operators $\hat{K}_r$:

\begin{equation}
\hat{\varrho}_{\mbox{\scriptsize out}}(t) = \sum_{r}\hat{K}_r(t)
\hat{\varrho}_{\mbox{\scriptsize in}} \hat{K}^\dagger_r(t),
\end{equation}
where \cite{ref-Daffer}:

\begin{widetext}
\begin{eqnarray}
\label{eq-Kraus} \hat{K}_0 &=&
\frac{1}{2}\hat{\openone}\sqrt{1+\Lambda_x+\Lambda_y+\Lambda_z}
= \hat{\openone}\sqrt{\int\mbox{d}^3r\,p({\bf r})\cos^2 rt}, \nonumber \\ \nonumber \\
\hat{K}_1
&=&\frac{1}{2}\hat{\sigma}_x\sqrt{1+\Lambda_x-\Lambda_y-\Lambda_z}
= \hat{\sigma}_x\sqrt{\int\mbox{d}^3r\,p({\bf r})\frac{x^2}{r^2}\sin^2 rt},\nonumber \\ \nonumber \\
\hat{K}_2&=&\frac{1}{2}\hat{\sigma}_y\sqrt{1-\Lambda_x+\Lambda_y-\Lambda_z}
= \hat{\sigma}_y\sqrt{\int\mbox{d}^3r\,p({\bf r})\frac{y^2}{r^2}\sin^2 rt},\nonumber \\ \nonumber \\
\hat{K}_3 &=&
\frac{1}{2}\hat{\sigma}_z\sqrt{1-\Lambda_x-\Lambda_y+\Lambda_z} =
\hat{\sigma}_z\sqrt{\int\mbox{d}^3r\,p({\bf r})
\frac{z^2}{r^2}\sin^2 rt}.
\end{eqnarray}
\end{widetext}

These exact expressions for the Kraus operators, are the main
result of our investigations. Before we discuss various
statistical models, we  note that an unbiased incoherent addition
of bit-error channels in most cases is not justified. The formula
above shows that the time-dependent $\Lambda_i(t)$ functions
couple  in a highly nontrivial way  the three channels. The
simplified expression (\ref{unbchannel}), does not reflect this
complicated entanglement between various bit-error channels.

\section{Examples of a noise\label{sec-examples}}

\subsection{ Markov noise}

Consider a simple case of a completely positive map determined by
a Lorentzian probability distribution

\begin{eqnarray}
p({\bf r}) &=&
\frac{1}{3\pi}\left(\frac{\Gamma/2}{x^2+\Gamma^2/4}\delta(y)\delta(z)\right.\nonumber \\ \nonumber \\
& &\left.+ \delta(x)\frac{\Gamma/2}{y^2+\Gamma^2/4}\delta(z)+
\delta(x)\delta(y)\frac{\Gamma/2}{z^2+\Gamma^2/4} \right)\nonumber \\
\end{eqnarray}
characterized by a width $\Gamma$. The corresponding Kraus
operators (\ref{eq-Kraus}) in this case read

\begin{eqnarray}
\hat{K}_0 &=& \hat{\openone}\sqrt{\frac{1+\exp(-\Gamma t)}{2}},\nonumber \\
\hat{K}_1 &=& \hat{\sigma}_x\sqrt{\frac{1-\exp(-\Gamma t)}{2}},\nonumber \\
\hat{K}_2 &=& \hat{\sigma}_y\sqrt{\frac{1-\exp(-\Gamma t)}{2}},\nonumber \\
\hat{K}_3 &=& \hat{\sigma}_z\sqrt{\frac{1-\exp(-\Gamma t)}{2}}.
\end{eqnarray}
The resulting dynamics of  such a channel is:
\begin{eqnarray}
\hat{\varrho}(t) &=& \frac{1+\exp(-\Gamma t)}{2}\hat{\varrho}(0) +
\frac{1-\exp(-\Gamma t)}{6}\nonumber \\ \nonumber \\
& &\times\left[\hat{\sigma}_x\hat{\varrho}(0)\hat{\sigma}_x +
\hat{\sigma}_y\hat{\varrho}(0)\hat{\sigma}_y +
\hat{\sigma}_z\hat{\varrho}(0)\hat{\sigma}_z \right].
\end{eqnarray}
Let us note that this expression is equivalent to
Eq.~(\ref{unbchannel}) if the probability is time-dependent i.e.,
\begin{equation}
 p(t)= \frac{1-\exp(-\Gamma t)}{2}.
\end{equation}
In the steady state $p(\infty) =\frac{1}{2}$, the quantum channel
reduces to a very simple expression \cite{ref-Preskill}:

\begin{equation}
\label{eq-preskill} \hat{\varrho}_{\mbox{\scriptsize out}} =
\frac{1}{2}\left(\hat{\varrho}_{\mbox{\scriptsize
in}}+\frac{1}{3}\hat{\sigma}_x \hat{\varrho}_{\mbox{\scriptsize
in}} \hat{\sigma}_x +\frac{1}{3}\hat{\sigma}_y
\hat{\varrho}_{\mbox{\scriptsize in}} \hat{\sigma}_y
+\frac{1}{3}\hat{\sigma}_z \hat{\varrho}_{\mbox{\scriptsize in}}
\hat{\sigma}_z\right).
\end{equation}

One can easily check, that the infinitesimal time evolution of the
last three Kraus operators is: $\hat{K}_i(\Delta t) \simeq
\sqrt{\Delta t}\, \hat{\sigma}_i$. This behavior is typical for a
diffusion process and consequently the state evolution  clearly
obeys the Markov no-memory property, and as a consequence has the
form of the Lindblad equation (\ref{lindblad}):

\begin{eqnarray}
\frac{\mbox{d}\hat{\varrho}}{\mbox{d}t} &=& \mathcal{L}
\hat{\varrho}(t)\nonumber\\
&=&-\frac{\Gamma}{2}\left[\hat{\varrho}-\frac{1}{3}\left(
\hat{\sigma}_x\hat{\varrho}\hat{\sigma}_x +
\hat{\sigma}_y\hat{\varrho}\hat{\sigma}_y +
\hat{\sigma}_z\hat{\varrho}\hat{\sigma}_z\right) \right].
\end{eqnarray}

\subsection{Telegraphic non-Markov noise}

Now, let us assume, that the noise introduced to the system is a
random telegraphic noise \cite{vankampen} , so that the
disturbance of the qubit induced by the environment is discrete,
and jumps between two values $\pm a$. For concreteness we consider
the following probability distribution:

\begin{equation}
p({\bf r})=\frac{1}{2}\left[\delta(x-a)+\delta(x+a)\right]
\delta(y)\delta(z).
\end{equation}
In this case the Kraus operators (\ref{eq-Kraus}) equal:

\begin{eqnarray}
\hat{K}_0 &=& \hat{\openone}\sqrt{\cos^2 at},\nonumber \\
\hat{K}_1 &=& \hat{\sigma_x}\sqrt{\sin^2 at},\nonumber \\
\hat{K}_2 &=& \hat{K}_3=0
\end{eqnarray}
and the disturbed qubit at instant $t$ is in the state:

\begin{equation}
\hat{\varrho}(t)=\cos^2 at \,\hat{\varrho}(0) + \sin^2 at\,
\hat{\sigma}_x \hat{\varrho}(0) \hat{\sigma}_x.
\end{equation}
The periodic result is very straightforward, however it reveals
something interesting. Although our model is quite simple, it
leads to non-trivial dynamics, which becomes apparent when we
analyze the evolution of the density operator
$\frac{\mbox{\scriptsize d}{\hat{\varrho}}}{\mbox{\scriptsize
d}t}$. One can easily find that the time-evolution is given by a
non-local in time Lindblad equation:

\begin{equation}
\label{eq-telegrdynamics}
\frac{\mbox{d}\hat{\varrho}(t)}{\mbox{d}t} =
-\frac{a^2}{2}\int_0^t
\mbox{d}s\left[\hat{\varrho}(s)-\hat{\sigma}_x \hat{\varrho}(s)
\hat{\sigma}_x \right].
\end{equation}
This shows that the time dynamics of $\hat{\varrho}(t)$ at the
given instant $t$ depends not only on the state at this instant,
but also on the state at all earlier times. This behavior can be
seen already from the form of the Kraus operator, which for the
infinitesimal time  evolves as: $\hat{K}_1(\Delta t) \simeq \Delta
t\, \hat{\sigma}_x$ and such an evolution characterizes non-Markov
processes with zero-bandwidth \cite{ref-Daffer}.

From this example we conclude that our model in general does not
obey the ``no-memory" approximation and the evolution of the state
is non-Markovian. One could also think of studying
multi-dimensional telegraphic noise, however in this case the
analysis becomes much more complicated and there is no simple,
linear integral kernel as the one in
Eq.~(\ref{eq-telegrdynamics}).

\begin{figure}
\begin{center}
\epsfig{file=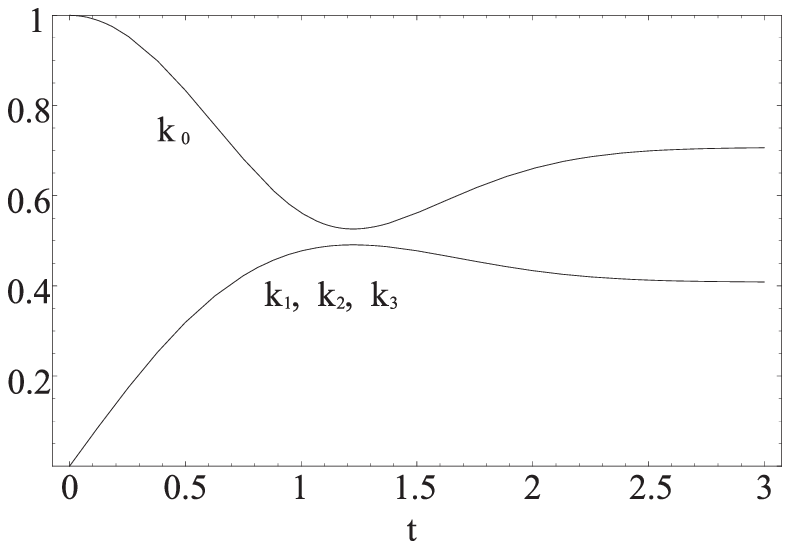} \epsfig{file=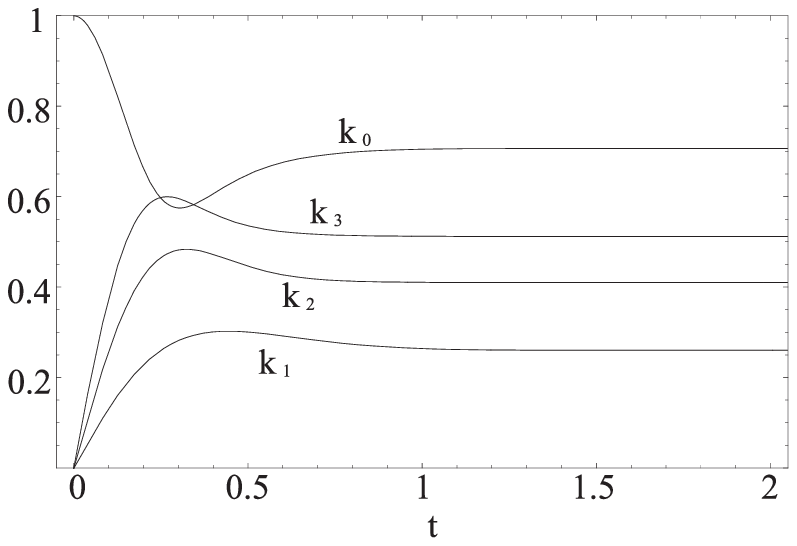}
\end{center}
\caption{\label{fig-Kraus} Kraus operator's coefficients $k_i$
(defined via the relation: $\hat{K}_i=k_i\hat{\sigma}_i$) for a
Gaussian probability distribution $p({\bf r})=\frac{1}{\sqrt{\pi^3
d_x^2 d_y^2 d_z^2}} \exp\left(-\frac{x^2}{d_x^2}
-\frac{y^2}{d_y^2} -\frac{z^2}{d_z^2}\right)$, where on the upper
figure $d_x=d_y=d_z=1$ and on the lower $d_x=1$, $d_y=2$,
$d_z=3$.}
\end{figure}

\subsection{Gaussian noise}
Although in general the expressions (\ref{eq-Kraus}) are not
analytically integrable, one can find explicitly the Kraus
operators in the asymptotic steady-state limit
$t\rightarrow\infty$. In this limit the square of rapidly
oscillating trigonometric functions appearing in the integrals
(\ref{eq-Kraus}) can be approximated by their average value
$\frac{1}{2}$. In the simplest case of an arbitrary, spherically
symmetric probability distribution $p(r)$ the coefficients $k_i$
of the Kraus operators (defined via the relation
$\hat{K}_i=k_i\hat{\sigma}_i$) equal $k_0=\frac{1}{\sqrt{2}}$,
$k_1=k_2=k_3=\frac{1}{\sqrt{6}}$ and consequently the output
quantum state reads as in Eq.~(\ref{eq-preskill}).

This result reproduces  the steady-state  Markov limit  justifying
its validity. However it is valid only when the probability
distribution $p({\bf r})$ is spherically symmetric i.e., the three
channels are unbiased.  In general the input state evolves to a
different limit.

In the Figure \ref{fig-Kraus} we have shown the numerically
calculated evolution of the Kraus operator's coefficients for a
Gaussian probability distribution

\begin{equation}
p({\bf r})=\frac{1}{\sqrt{\pi^3 d_x^2 d_y^2 d_z^2}}
\exp\left(-\frac{x^2}{d_x^2} -\frac{y^2}{d_y^2}
-\frac{z^2}{d_z^2}\right)
\end{equation}
for the following two cases: when the probability $p(r)$ is
spherically symmetric with $d_x=d_y=d_z=1$ (upper plot) and for
the asymmetric distribution with $d_x=1$, $d_y=2$, $d_z=3$ (lower
plot). It is seen that after some characteristic time, the state
becomes stationary, however the limit depends on the
characteristics of the probability distribution $p({\bf r})$. The
example discussed above provides an illustration of a unbiased and
biased Gaussian depolarization channels.

\section{Fidelity for mixed input states\label{sec-fidelity}}

In order to judge the quality of a communication channel and the
role  of the introduced noise one needs a tool to investigate the
state disturbance during the transmission. To quantify the
 influence of the external noise onto the transmitted
quantum state we use an entanglement fidelity measure defined as
the following overlap between the input and output density matrix
\cite{ref-Nielsen}:

\begin{equation}
\mathcal{F}(\hat{\varrho}_{\mbox{\scriptsize in}},
\hat{\varrho}_{\mbox{\scriptsize out}}) = \left(\mbox{Tr}\left\{
\sqrt{\hat{\varrho}^{\frac{1}{2}}_{\mbox{\scriptsize in}}\,
\hat{\varrho}_{\mbox{\scriptsize out}}\,
\hat{\varrho}^{\frac{1}{2}}_{\mbox{\scriptsize
in}}}\right\}\right)^2.
\end{equation}
This fidelity is in general  very difficult or impossible to
calculate. For an arbitrary input state of a single qubit given by
Eq.~(\ref{initial}), and with an arbitrary spherically symmetric
probability distribution $p(r)$ characterizing the external noise
this fidelity can be calculated exactly and is equal to:

\begin{equation}
\mathcal{F}(\hat{\varrho}_{\mbox{\scriptsize in}},
\hat{\varrho}_{\mbox{\scriptsize out}}) = \frac{1}{2}\left( \xi +
\sqrt{\chi(1-a^2)} \right),
\end{equation}
where $\xi=1+a_x^2\Lambda_x+a_y^2\Lambda_y+a_z^2\Lambda_z$ and
$\chi = 1-a_x^2\Lambda^2_x-a_y^2\Lambda^2_y-a_z^2\Lambda^2_z$. For
pure states ($a=1$) the formula simplifies to
$\mathcal{F}=\frac{\xi}{2}$. On the other hand it is not very
surprising, that the maximally mixed state ($a=0$) remains
unchanged under the influence of the noise, while the
communication fidelity decreases with increasing purity of the
input state.

Using the same approach one may study also the evolution of
multidimensional systems. Of course the general expressions become
very complicated, even when we consider a two-qubit Hilbert space,
however it is possible to find some compact solutions for special
cases of pure states.

Consider an arbitrary two-qubit initial pure state:

\begin{equation}
 |\Psi_{\mbox{\scriptsize in}}\rangle =
a|\!\updownarrow\updownarrow\rangle
+b|\!\updownarrow\leftrightarrow\rangle
+c|\!\leftrightarrow\updownarrow\nolinebreak\rangle
+d|\!\leftrightarrow\leftrightarrow\nolinebreak\rangle.
\end{equation}
Using the same approach as above, one can calculate that for the
independent disturbance of each mode with the same type of noise
characterized by the spherically symmetric probability
distribution $p(r)$ the fidelity of the transformation is:

\begin{figure}
\begin{center}
\epsfig{file=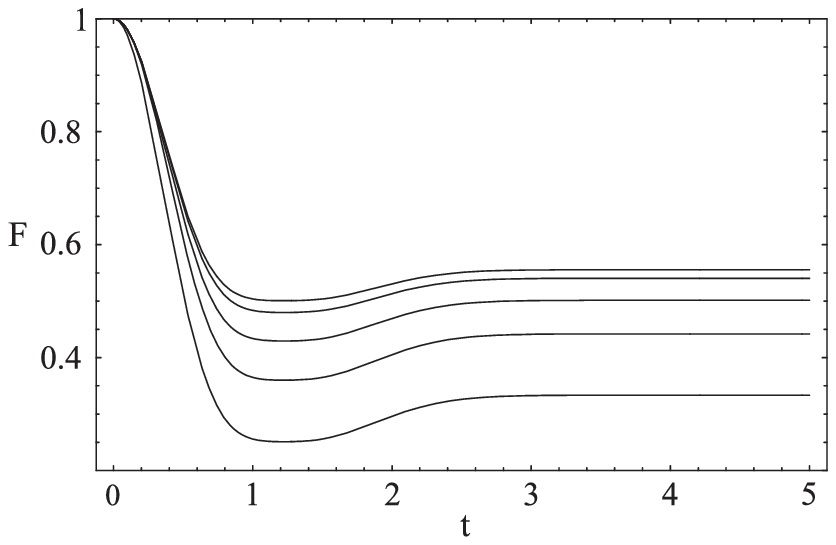}
\end{center}
\caption{\label{fig-PartlyMixed} Fidelity measure as a function of
time $t$ for the input state $\hat{\varrho}_{\mbox{\scriptsize
in}}=\frac{1}{2}\left( |\!\updownarrow\updownarrow\rangle
\langle\,\updownarrow\updownarrow\!| +
|\!\leftrightarrow\leftrightarrow\rangle
\langle\,\leftrightarrow\leftrightarrow\!| \right)
+\frac{m}{2}\left( |\!\updownarrow\updownarrow\rangle
\langle\,\leftrightarrow\leftrightarrow\!| +
|\!\leftrightarrow\leftrightarrow\rangle
\langle\,\updownarrow\updownarrow\!| \right)$ for several values
of the parameter $m$. From bottom to the top, respectively: $m=1$,
$m=0.9$, $m=0.7$, $m=0.4$, $m=0$.}
\end{figure}

\begin{eqnarray}
\mathcal{F}(|\Psi\rangle_{\mbox{\scriptsize in}},
\hat{\varrho}_{\mbox{\scriptsize out}}) &=& \left(\mbox{Tr}\left\{
\sqrt{ |\Psi_{\mbox{\scriptsize in}}
\rangle\langle\Psi_{\mbox{\scriptsize in}}|
\hat{\varrho}_{\mbox{\scriptsize out}} |\Psi_{\mbox{\scriptsize
in}} \rangle\langle\Psi_{\mbox{\scriptsize in}}| }
\right\}\right)^2 \nonumber \\ \nonumber \\
&=& \langle\Psi_{\mbox{\scriptsize in}}|
\hat{\varrho}_{\mbox{\scriptsize out}} |\Psi_{\mbox{\scriptsize
in}} \rangle \nonumber \\ \nonumber \\
&=& \left(\frac{1+\Lambda}{2}\right)^2-4\Lambda
\frac{1-\Lambda}{2}|bc-ad|^2.
\end{eqnarray}
What is interesting in the above result is that the fidelity is
the highest for separable states (for example for $a=b=0$ and any
$c, d$) and it drops down when the input state becomes more
entangled.

Another compact result can be found for the following two-mode
mixed input state:

\begin{eqnarray}
\hat{\varrho}_{\mbox{\scriptsize in}} &=&\frac{1}{2}\left(
|\!\updownarrow\updownarrow\rangle\langle\,\updownarrow\updownarrow\!|+
|\!\leftrightarrow\leftrightarrow\rangle\langle\,\leftrightarrow\leftrightarrow\!|
\right) \nonumber \\ \nonumber \\
& &+ \frac{m}{2}\left( |\!\updownarrow\updownarrow\rangle
\langle\,\leftrightarrow\leftrightarrow\!| +
|\!\leftrightarrow\leftrightarrow\rangle
\langle\,\updownarrow\updownarrow\!| \right).
\end{eqnarray}
With a similar analysis one obtains the fidelity measure given by:

\begin{eqnarray}
F\left(\hat{\varrho}_{\mbox{\scriptsize
in}},\hat{\varrho}_{\mbox{\scriptsize out}}\right) &=& \frac{1}{4}
\left[1+\Lambda_z^2+m^2\left(\Lambda_x^2+\Lambda_y^2\right)
+\sqrt{1-m^2} \right.\nonumber \\ \nonumber \\
& &\left. \times\sqrt{\left(1+\Lambda_z^2\right)^2
-m^2\left(\Lambda_x^2+\Lambda_y^2\right)^2}\right].
\end{eqnarray}
In the Figure \ref{fig-PartlyMixed} we have plotted the dynamics
of fidelity for several parameters $m$ and the unbiased Gaussian
probability distribution $p(r) = \pi^{-\frac{3}{2}} e^{-r^2}$. We
find a not very surprising result, that the transformation
fidelity is a decreasing function of the purity of the input
state. For $m=0$ (which of course does not yet correspond to the
maximally mixed state) the fidelity is the highest, while for the
pure state ($m=1$) the fidelity is the lowest.

\section{Conclusions\label{sec-conclusions}}
Uncontrolled interaction between the environment and the
transmitted quantum state can essentially affect the state and in
consequence lower the communication capacity of an information
channel. Several ideas has been put forth to overcome this
problem. One of the most promising is the use of so-called
decoherence-free subspaces \cite{ref-Kwiat,ref-Banaszek}. This
idea can be applied when the noise present in the system is
correlated between consecutive uses of the communication channel
\cite{ref-Ball}, however this is not always possible and therefore
one needs a careful study of the properties of various types of
noise and their influence on the quantum state.

In this paper we have introduced a dynamical model of interaction
between the quantum state and its environment and shown that
although based on simple assumptions, it leads to non-trivial
solutions. Using the model we have analyzed properties of
zero-bandwidth noise with Lorentzian, telegraphic and Gaussian
distributions and have shown that only the first of them obeys the
Markov property, while the others exhibit memory effects and are
non-Markovian. Our approach allowed us to solve a simplified
version of a general problem when the noise is an arbitrary
time-dependent stochastic process, whose solution is not known. We
have calculated transformation fidelities for a collection of
input states and analyzed their dynamics according to their
entanglement or purity.

\begin{acknowledgments}
This work was partially supported by a KBN grant No. 2PO3B 02123.
A.D. thanks The Foundation for Polish Science for the support with
the Annual Stipend for Young Scientists.

\end{acknowledgments}

\end{document}